\documentclass[aps,pra,twocolumn,10pt]{revtex4-1}

\usepackage{amsmath,amssymb,bm}
\usepackage{dsfont}
\usepackage{graphics}

\newcommand{\tr}{\text{tr}}
\newcommand{\ket}[1]{\left|{#1}\right\rangle}
\newcommand{\bra}[1]{\left\langle{#1}\right|}
\newcommand{\Dcirc}{{\cal D}^{\circ}}

\begin{document}

\title{Distortions produced in optical homodyne tomography}
\author{Filippus S. \surname{Roux}}
\email{froux@nmisa.org}
\affiliation{National Metrology Institute of South Africa, Meiring Naud{\'e} Road, Brummeria 0040, Pretoria, South Africa}

\begin{abstract}
An analysis of the homodyne tomography process that is often used to determine the Wigner functions of quantum optical states is performed to consider the effects of the spatiotemporal degrees of freedom. The homodyne tomography process removes those parts of the input state that are not associated with the mode of the local oscillator by tracing out those degrees of freedom. Using a functional approach to incorporate all the spatiotemporal degrees of freedom, we find that this reduction in the degrees of freedom introduces distortions in the observed Wigner function. The analysis also shows how the homodyne tomography process introduces a resolution that depends on the strength of the local oscillator. As examples, we consider coherent states, Fock states and squeezed vacuum states.
\end{abstract}

\maketitle

\section{\label{intro}Introduction}

Homodyne tomography \cite{lvovsky0} is widely used to determine the Wigner functions of quantum optical states in terms of their particle-number degrees of freedom, pertaining to specific spatiotemporal modes. It has been used to measure the Wigner functions of
squeezed vacuum states \cite{smithey2,ohtsqu}, Fock states \cite{lvovfock,ohtfock1,ohtfock2}, photon added states \cite{zav1,zav3}, and many others. The quality of experimentally prepared special quantum states, used as resources in quantum information systems, is determined with the aid of homodyne tomography. However, it begs the question of the quality of the homodyne tomography process itself.

Various aspects of the homodyne tomography process have been investigated \cite{kvogel,leonhardt,kuhn}, including the temporal effects \cite{ohtbandw}, and the efficiency and noise of detector systems \cite{raymerdc,ohtnoise}. Mathematical and statistical methods with which Wigner functions are constructed from the measured data have been improved significantly over time.

These analyses generally assume that the measurements from which the Wigner functions of quantum states are constructed are restricted to the part of the Hilbert space associated with the mode of the local oscillator, regardless of the complexity of this mode. In free space, a quantum optical state contains an infinite number of spatiotemporal degrees of freedom in addition to its particle-number degrees of freedom. It is not possible the measure all these degrees of freedom in a tomography process. Some form of dimensional reduction is inevitable in any such measurement process. Homodyne tomography imposes this dimensional reduction primarily through an overlap by the mode of the local oscillator, but the detector system can also have an effect on the dimensional reduction process. All the unobserved degrees of freedom of the state are traced out.

Here, the intrinsic fidelity of the homodyne tomography process is investigated. We use a Wigner functional approach \cite{stquad,nosemi,paramp}, allowing us to incorporate all the spatiotemporal degrees of freedom in the analysis. In addition, generating functions \cite{toolbox} are used extensively.

\section{Optical homodyning}

\begin{figure}[ht]
\centerline{\includegraphics{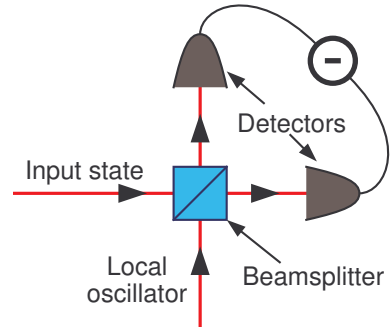}}
\caption{Diagram of the homodyne tomography system.}
\label{homosyst}
\end{figure}

There are different versions of the homodyning system that has been developed since its inception (see \cite{lvovsky0} and references therein). They include heterodyning and double homodyning systems \cite{doublehomod0,doublehomod}. However, we consider the basic homodyning system here, as depicted in Fig.~\ref{homosyst}. The local oscillator is a coherent state with a mode that is parameterized in terms of a spectral function. The input state is mixed with the local oscillator via a 50:50 beamsplitter. The light from both output ports of the beamsplitter are sent to identical detectors. The intensities registered by these detectors are subtracted from each other and then binned to provide a photon number probability distribution.

Formally, we represent the quantum optical state to be measured in terms of all its degrees of freedom by using a Wigner functional $W[\alpha]$, defined on a functional phase space where $\alpha(\mathbf{k})$ is the functional's field variable (itself a spectral function of the wavevector $\mathbf{k}$). The measurement process induces the reduction of the Wigner functional to a Wigner {\em function} $W(\alpha_0)$, where $\alpha_0$ is a complex variable defined on a two-dimensional subspace of the full infinite dimensional functional phase space. Naively, this reduction process implies that the field variable of the Wigner functional is replace by $\alpha(\mathbf{k})\rightarrow \alpha_0\Gamma(\mathbf{k})$ where $\Gamma(\mathbf{k})$ is the normalized mode of the local oscillator, represented as an angular spectrum, and that all the other degrees of freedom are simply discarded by the trace process. It turns out that the actual dimensional reduction process associated with homodyne tomography is in general more complicated than this naive view.

\subsection{Cross-correlation function}

When homodyning measurements are used to perform quantum state tomography, it is necessary to measure the photon-number statistics of the difference in intensity. Instead of the number operator measuring the average intensity, we need to use the projection operators for $n$ photons for the analysis. Separate projection operators are used for the respective detectors, leading to two separate photon-number distributions for the two detectors. The difference between the measurements corresponds to the convolution of the distribution at one detector with the mirror image of the distribution at the other detector, which is the same as the cross-correlation function of the two distributions.

Assume that $P_1(n)$ and $P_2(n)$ represent the probability distributions for detecting photons at the two respective detectors. The cross-correlation of the two distributions is then given by
\begin{equation}
R(m) = \sum_{n=0}^{\infty} P_1(n) P_2(n+m) ,
\end{equation}
where $m$ can be any signed integer. The requirement that $n+m>0$ is maintained by the distributions themselves, being zero for negative arguments. A generating function for $R(m)$ is obtained by multiplying it by $K^m$ and summing over $m$:
\begin{align}
\mathcal{R}(K) = & \sum_{m=-\infty}^{\infty} K^m R(m) \nonumber \\
 = & \sum_{n=0}^{\infty} \sum_{m=-\infty}^{\infty} K^m P_1(n) P_2(n+m) .
\label{spreikor0}
\end{align}
Then we redefine $m\rightarrow p-n$ to get
\begin{equation}
\mathcal{R}(K) = \sum_{n,p=0}^{\infty} K^{p-n} P_1(n) P_2(p)
= \mathcal{P}_1(K) \mathcal{P}_2(K^{-1}) ,
\label{spreikor}
\end{equation}
where we enforced the positivity of both arguments to obtain summations that start from $0$, and where we introduced the generating functions for the original distributions, given by
\begin{equation}
\mathcal{P}_{1,2}(K) = \sum_{n=0}^{\infty} K^n P_{1,2}(n) .
\end{equation}
As such, the generating function in Eq.~(\ref{spreikor}) for the cross-correlation of the two distributions is given in terms of the generating functions of the respective distributions.

Here, we treated $P_1(n)$ and $P_2(n)$ as being statistically independent distributions. However, they are measured at the same time and the correlation is based on these simultaneous measurements. Therefore, the cross-correlation should be represented as a combined operator that is traced with the state to determine the cross-correlated distribution. Based on Eq.~(\ref{spreikor}), a generating function for such operators is of the form
\begin{equation}
\hat{\mathcal{R}}(K) = \sum_{n=0}^{\infty} \sum_{p=0}^{\infty} K^{p-n} \hat{P}_n^{(1)} \hat{P}_p^{(2)}
 = \hat{\mathcal{P}}_1(K) \hat{\mathcal{P}}_2(K^{-1}) ,
\label{spreikorop}
\end{equation}
where $\hat{\mathcal{P}}_1(K)$ and $\hat{\mathcal{P}}_2(K)$ are the generating functions for the $n$-photon projection operators associated with the respective detectors.

\subsection{Wigner functionals}

The generating function for the Wigner functionals of the $n$-photon projection operators is given by
\begin{equation}
\mathcal{W}(K) = \left(\frac{2}{1+K}\right)^{\lambda} \exp\left(-2\frac{1-K}{1+K}\alpha^*\diamond D\diamond\alpha\right) ,
\label{fotnumgen}
\end{equation}
where the $\diamond$-contraction is represents an integration over the wavevectors,
\begin{equation}
\alpha^*\diamond D\diamond\alpha \equiv \int \alpha^*(\mathbf{k}) D(\mathbf{k},\mathbf{k}') \alpha(\mathbf{k}')\ \frac{d^2 k d\omega}{(2\pi)^3}\ \frac{d^2 k' d\omega'}{(2\pi)^3} ,
\label{binnespek}
\end{equation}
$D(\mathbf{k},\mathbf{k}')$ is an idempotent kernel that represents the detection process, and $\lambda=\tr\{D\}$ counts the number of degrees of freedom that can pass through the detector. The detector kernel represents a photon-number resolving detector. However, it usually also incorporates spatiotemporal degrees of freedom imposing restrictions on the modes that can be detected.

The Wigner functional of the combined operator of the two detectors is represented by
\begin{equation}
\mathcal{W}_{\hat{R}}
= \mathcal{M}_K^{\lambda} \exp\left[-2 \mathcal{J}_K(\alpha^*\diamond D\diamond\alpha -\beta^*\diamond D\diamond\beta)\right] ,
\label{spreikorwig1}
\end{equation}
where $\alpha$ and $\beta$ are the field variables associated with the respective detectors, $D$ is the same detector kernel for both detectors (they need to be identical for successful homodyning), $K$ is the generating parameter, and
\begin{align}
\begin{split}
\mathcal{M}_K = & \frac{4K}{(1+K)^2} , \\
\mathcal{J}_K = & \frac{1-K}{1+K} .
\end{split}
\label{auxfunc}
\end{align}

\subsection{Beamsplitter}

The cross-correlation operator is traced with the state that is obtained after the beamsplitter. The measurements can therefore be represented by
\begin{equation}
\langle \hat{R} \rangle = \tr\left\{\hat{U}_{\text{BS}} (\hat{\rho}_{\text{in}} \otimes \hat{\rho}_{\text{lo}}) \hat{U}_{\text{BS}}^{\dag} \hat{R} \right\} ,
\end{equation}
where $\hat{\rho}_{\text{in}}$ and $\hat{\rho}_{\text{lo}}$ are the density operators for the input state and the local oscillator, respectively, and $\hat{U}_{\text{BS}}$ is the unitary operator for the beamsplitter. Combined with the beamsplitters' unitary operators, the detection operator becomes
\begin{equation}
\hat{R}' = \hat{U}_{\text{BS}}^{\dag} \hat{R}\hat{U}_{\text{BS}} .
\label{defuru}
\end{equation}
The unitary transformation for a 50:50 beamsplitter applied to the Wigner functional of a state is represented by a transformation of the field variables in the argument of the Wigner functional, given by
\begin{align}
\begin{split}
\alpha(\mathbf{k}) \rightarrow & \frac{1}{\sqrt{2}} [\alpha(\mathbf{k}) + i \beta(\mathbf{k})] \\
\beta(\mathbf{k}) \rightarrow & \frac{1}{\sqrt{2}} [\beta(\mathbf{k}) + i \alpha(\mathbf{k})] .
\end{split}
\label{wigbstra}
\end{align}
However, since the unitary operators appear in the opposite order in Eq.~(\ref{defuru}), we need to apply the inverse transformations to the Wigner functional in Eq.~(\ref{spreikorwig1}), and thereby obtain
\begin{equation}
\mathcal{W}_{\hat{R}}'
= \mathcal{M}_K^{\lambda} \exp\left[i 2 \mathcal{J}_K(\beta^*\diamond D\diamond\alpha -\alpha^*\diamond D\diamond\beta)\right] .
\label{spreikorbs0}
\end{equation}

\subsection{Local oscillator}

The generating function in Eq.~(\ref{spreikorbs0}) is multiplied with the Wigner functional for the local oscillator before the beamsplitter and the result is traced over the local oscillator degrees of freedom. The result is represented in terms of operators as
\begin{equation}
\hat{H} = \tr_{\text{lo}}\left\{\hat{\rho}_{\text{lo}} \hat{U}_{\text{BS}}^{\dag} \hat{R} \hat{U}_{\text{BS}} \right\} .
\end{equation}
In terms of the Wigner functionals, the trace is performed by evaluating the functional integration over $\beta$, which is the field variable associated with the local oscillator. The trace produces the Wigner functional of the operator that is used to produce the distribution obtained from the homodyne process. It reads
\begin{align}
\mathcal{W}_{\hat{H}} [\alpha] = & \int W_{\text{lo}}[\beta] \mathcal{W}_{\hat{R}}'[\alpha,\beta]\ \Dcirc[\beta] \nonumber \\
= & \mathcal{M}_K^{\lambda}
\exp\left[i 2\mathcal{J}_K (\gamma^*\diamond D\diamond\alpha -\alpha^*\diamond D\diamond\gamma) \right. \nonumber \\
& \left. +2\mathcal{J}_K^2 \alpha^*\diamond D\diamond\alpha\right] ,
\label{spreikorbs}
\end{align}
where $W_{\text{lo}}[\beta]$ is the Wigner functional of a coherent state for the local oscillator parameterized by $\gamma(\mathbf{k})$, which is the parameter function (mode function) of the local oscillator, and $\Dcirc[\beta]$ is the functional integration measure.

The exponent in Eq.~(\ref{spreikorbs}) contains the terms that combine into the contraction of the local oscillator mode with a real valued field variable (the quadrature variable) along a direction determined by the phase of the local oscillator mode. The exponent also contains a term that is independent of the local oscillator mode, and which is responsible for some of the distortions.

\section{Probability distribution}

The generating function for the distribution produced by the homodyning process is obtained by multiplying the Wigner functional of the state $W[\alpha]$ by Eq.~(\ref{spreikorbs}) and computing the trace of the product:
\begin{equation}
\mathcal{W}_{\text{H}}(K) = \int W[\alpha] \mathcal{W}_{\hat{H}}[\alpha]\ \Dcirc[\alpha] .
\label{spreikorbs1}
\end{equation}
Following the inverse Radon transform approach \cite{smithey2} to obtain the observed Wigner function from the homodyning experimental results, we need to extract the probability distribution. The generating function for the distribution, represented by Eq.~(\ref{spreikorbs1}), is the same generating function given in Eq.~(\ref{spreikor0}). Hence,
\begin{equation}
\mathcal{W}_{\text{H}}(K) = \mathcal{R}(K) = \sum_{m=-\infty}^{\infty} K^m R(m) ,
\label{homogen0}
\end{equation}
where $R(m)$ is the probability distribution for the cross-correlation. Since the index $m$ also runs over negative integers, we cannot extract individual terms with the aid of derivatives as is often done with generating functions. Instead, the individual probabilities are extracted with the aid of an auxiliary integral for the Kronecker delta,
\begin{equation}
\frac{1}{2\pi} \int_{-\pi}^{\pi} \exp[i(m-n)\phi]\ \text{d}\phi = \delta_{m,n} .
\label{genkron}
\end{equation}
It implies that the probability distribution for the cross-correlation is extracted from its generating function by
\begin{align}
R(n) = & \sum_{m=-\infty}^{\infty} R(m) \frac{1}{2\pi} \int_{-\pi}^{\pi} \exp[i(m-n)\phi]\ \text{d}\phi \nonumber \\
 = & \frac{1}{2\pi} \int_{-\pi}^{\pi} \exp(-i n\phi) \mathcal{W}_{\text{H}}(e^{i\phi})\ \text{d}\phi .
\label{genhomo}
\end{align}
The expression in Eq.~(\ref{spreikorbs}) is substitute into Eq.~(\ref{spreikorbs1}), which is then substituted into Eq.~(\ref{genhomo}). When we replace $K=\exp(i\phi)$ in $\mathcal{J}_K$ and $\mathcal{M}_K$, they become
\begin{align}
\begin{split}
\mathcal{J}_K \rightarrow & -i \tan(\tfrac{1}{2}\phi) , \\
\mathcal{M}_K \rightarrow & \frac{1}{\cos^2(\tfrac{1}{2}\phi)} .
\end{split}
\end{align}
The expression for the distribution thus becomes
\begin{align}
R(n) = & \frac{1}{2\pi} \int_{-\pi}^{\pi} \frac{\exp(-i n\phi)}{\cos^{2\lambda}(\tfrac{1}{2}\phi)} \int W[\alpha] \nonumber \\
& \times \exp\left[ 2\tan(\tfrac{1}{2}\phi) (\gamma^*\diamond D\diamond\alpha-\alpha^*\diamond D\diamond\gamma) \right.\nonumber \\
& \left. -2\tan^2(\tfrac{1}{2}\phi) \alpha^*\diamond D\diamond\alpha\right]\ \Dcirc[\alpha]\ \text{d}\phi .
\label{genhomophi}
\end{align}

For convenience, the parameter function of the local oscillator is represented as $\gamma(\mathbf{k})=\gamma_0 \exp(i\theta) \Gamma(\mathbf{k})$, where
\begin{align}
\gamma_0 \equiv \|\gamma\| =  \sqrt{\gamma^*\diamond\gamma} ,
\end{align}
is the magnitude of the parameter function, $\Gamma(\mathbf{k})$ is a normalized spectral function, so that $\|\Gamma\|=1$, and $\theta$ is a variable phase. The distribution is now treated as a function of a continuous variable $x$. We define
\begin{align}
x= n\Delta x = \frac{n}{\gamma_0} ,
\end{align}
where we use the inverse of the magnitude of the local oscillator mode function to represent the small increment $\Delta x=\gamma_0^{-1}$. The distribution then becomes
\begin{align}
R(x,\theta) = & \frac{1}{2\pi} \int_{-\pi}^{\pi} \frac{\exp(-i x\gamma_0\phi)}{\cos^{2\lambda}(\tfrac{1}{2}\phi)} \int W[\alpha] \nonumber \\
& \times \exp\left[ 2\tan(\tfrac{1}{2}\phi) (\gamma^*\diamond D\diamond\alpha-\alpha^*\diamond D\diamond\gamma) \right.\nonumber \\
& \left. -2\tan^2(\tfrac{1}{2}\phi) \alpha^*\diamond D\diamond\alpha\right]\ \Dcirc[\alpha]\ \text{d}\phi ,
\label{genhomophi1}
\end{align}
where we show the probability distribution's dependence on the phase of the local oscillator $\theta$.

\section{Observed Wigner function}

To recover the {\em observed} Wigner function from the measured probability distribution, we perform two steps that implement the inverse Randon transform. The probability distribution in terms of $x$ is interpreted as a marginal distribution obtained from the partial integration of the Wigner functional, retaining only a one-dimensional variation along a direction determined by $\theta$. The result is a function and not a functional. In the first step, this marginal distribution is converted into a corresponding slice of the associated characteristic function via a Fourier transform
\begin{equation}
\chi(r,\theta) = \int R(x,\theta) \exp(i x r)\ \text{d} x ,
\label{genhomos}
\end{equation}
where $r$ and $\theta$ are treated as cylindrical coordinates, but with ranges given by $-\infty<r<\infty$ and $0\leq\theta\leq\pi$. When we substitute Eq.~(\ref{genhomophi1}) into Eq.~(\ref{genhomos}) and evaluate the integral over $x$, it produces a Dirac delta function
\begin{equation}
\int \exp(-i x\gamma_0\phi) \exp(i x r)\ \text{d} x = 2\pi \delta(\gamma_0\phi- r) .
\end{equation}
The integration over $\phi$ therefore replaces
\begin{equation}
\phi\rightarrow \frac{r}{\gamma_0}= r\Delta x .
\end{equation}
Hence, it imposes a boundary on the characteristic function. Since $-\pi<\phi<\pi$, it follows that $-\pi\gamma_0< r<\pi\gamma_0$. Provided that the characteristic function lies within this region, we can ignore the boundary. Otherwise the characteristic function would be clipped by the boundary. We'll assume that $\gamma_0$ is large enough that the characteristic function is contained inside this boundary.

In the second step, a symplectic Fourier transform is applied to the characteristic function to produce the observed Wigner function as a function of $q$ and $p$. It reads
\begin{equation}
W'(q,p) = \frac{1}{2\pi} \int \chi(\xi,\zeta) \exp\left(i q\xi-i p\zeta\right)\ \text{d}\zeta\ \text{d}\xi ,
\end{equation}
where $\xi$ and $\zeta$ are Cartesian coordinates, associated with the cylindrical coordinates $r$ and $\theta$, such that
\begin{equation}
 r^2 = \tfrac{1}{2} (\zeta^2+\xi^2) .
\end{equation}
The integrations over $x$ and $\phi$ in Eq.~(\ref{genhomos}) and Eq.~(\ref{genhomophi1}) then lead to
\begin{align}
W'(q,p) = & \mathcal{N} \int \frac{W[\alpha]}{\cos^{2\lambda}(\tfrac{1}{2} r\Delta x)} \nonumber \\
& \times \exp\left[-2\tan^2(\tfrac{1}{2} r\Delta x) \alpha^*\diamond D\diamond\alpha \right. \nonumber \\
& +2\tan(\tfrac{1}{2} r\Delta x)(\gamma^*\diamond D\diamond\alpha-\alpha^*\diamond D\diamond\gamma) \nonumber \\
& \left. +i q\xi-i p\zeta\right]\ \Dcirc[\alpha]\ \text{d}\zeta\ \text{d}\xi ,
\end{align}
where we introduce a normalization constant $\mathcal{N}$. For large enough $\gamma_0$ (small enough $\Delta x$),
\begin{align}
\begin{split}
\tan(\tfrac{1}{2} r\Delta x) = & \tfrac{1}{2} r\Delta x + O\left(r^3\Delta x^3\right) , \\
\cos(\tfrac{1}{2} r\Delta x) = & 1 + O\left(r^2\Delta x^2\right) .
\end{split}
\end{align}
If the characteristic function has a small enough size compare to the boundary, we can represent the observed Wigner function as
\begin{align}
W'(q,p) = & \mathcal{N} \int W[\alpha] \exp\left[-\tfrac{1}{2} r^2\Delta x^2 \alpha^*\diamond D\diamond\alpha \right. \nonumber \\
& + r\Delta x (\gamma^*\diamond D\diamond\alpha-\alpha^*\diamond D\diamond\gamma) \nonumber \\
& \left. +i q\xi-i p\zeta \right]\ \Dcirc[\alpha]\ \text{d}\zeta\ \text{d}\xi \nonumber \\
= & \mathcal{N} \int W[\alpha] \exp\left[-\tfrac{1}{4}(\zeta^2+\xi^2)\Delta x^2 \alpha^*\diamond D\diamond\alpha \right. \nonumber \\
& +\tfrac{1}{\sqrt{2}}(\zeta-i\xi)\Gamma^*\diamond D\diamond\alpha \nonumber \\
& -\tfrac{1}{\sqrt{2}}(\zeta+i\xi)\alpha^*\diamond D\diamond\Gamma \nonumber \\
& \left. +i q\xi-i p\zeta \right]\ \Dcirc[\alpha]\ \text{d}\zeta\ \text{d}\xi ,
\label{wnaw}
\end{align}
where we converted $r$, together with $\theta$ from within $\gamma$, into $\zeta$ and $\xi$ in the last expression.

Without the second-order term in the exponent, the integrations over $\zeta$ and $\xi$ would produce Dirac delta functions that would replace the contractions of $\alpha$ with $\Gamma$ via $D$ by $q$ and $p$. It would represent an ideal homodyning measurement process whereby the Wigner functional $W[\alpha]$ is converted to the observed Wigner function $W'(q,p)$, in which the functional integration replaces a two-dimensional subset of the degrees of freedom inside the Wigner functional by $q$ and $p$ and trace over all the other degrees of freedom.

The question is how to deal with the functional integration. For that, we need to consider the effect of the detector kernel in more detail.

\section{Detector kernel}

In general, the functional integration over $\alpha$ in Eq.~(\ref{wnaw}) cannot be evaluated, because $D$ is not invertible. It represents a projection operation that restricts the functional phase space to those functions that can be detected. Even if we discard the quadratic term, the remaining part of the argument in the exponent does not represent the entire functional phase space. The projection induced by the overlap with $\Gamma$ is in general even more restrictive than the projection associated with $D$. To evaluate the functional integration, we need to separate the integration into the subspaces defined by the projections imposed by $D$ and $\Gamma$.

Let's denote the total functional phase space by $\mathcal{A}$, the subspace onto which $D$ projects by $\mathcal{M}$, and the subspace associated with $\Gamma$ by $\mathcal{G}$. To be more precise, we state that for $\alpha\in\mathcal{M}$, we have $\alpha^*\diamond D\diamond\alpha\neq 0$, and for $\alpha\in\mathcal{G}$, we have $\alpha^*\diamond\Gamma\neq 0$. In the latter two cases, there are in general still parts of $\alpha$ that do not satisfy the requirements.

In the absurd case when $\mathcal{G}\cap\mathcal{M}=\emptyset$, which implies that $\Gamma^*\diamond D=D\diamond\Gamma=0$, (i.e., the detector cannot measure the mode of the local oscillator), the $\Gamma$-dependent terms in Eq.~(\ref{wnaw}) are zero, leaving us with
\begin{align}
W_0'(q,p) = & \mathcal{N} \int W[\alpha]
\exp\left[ -\tfrac{1}{4} (\zeta^2+\xi^2) \Delta x^2 \alpha^*\diamond D\diamond\alpha \right. \nonumber \\
& \left. +i q\xi-i p\zeta \right]\ \Dcirc[\alpha]\ \text{d}\zeta\ \text{d}\xi .
\label{absurd}
\end{align}
The result of the functional integration, which is simply the overlap of the Wigner functional of the state by a thermal states, is a rotationally symmetric function of $r$, peaked at the origin --- its amplitude at $r=0$ is the trace over the entire Wigner functional of the state. The Fourier transform of this function is also a rotationally symmetric function peaked at the origin. In other words, the absurd case produces a Wigner function reminiscent of that of a thermal state. Setting $\Delta x^2=0$, we get
\begin{align}
W_0'(q,p) = & \mathcal{N} \int W[\alpha] \exp(i q\xi-i p\zeta)\ \Dcirc[\alpha]\ \text{d}\zeta\ \text{d}\xi \nonumber \\
= & 4\pi^2\delta(q)\delta(p) .
\end{align}
Hence, for $\Delta x^2\neq 0$, the result is a narrow function at the origin with a width given by $\Delta x$.

Contrary to the absurd case, we shall assume that
\begin{equation}
\mathcal{G}\subset\mathcal{M}\subset\mathcal{A} .
\end{equation}
Then we can separate the phase space into three disjoint sets: $\mathcal{G}$, $\mathcal{M}_0$ and $\mathcal{A}_0$, where $\mathcal{M}_0$ is the part of $\mathcal{M}$ that excludes $\mathcal{G}$ and $\mathcal{A}_0$ is the part of $\mathcal{A}$ excluding $\mathcal{M}$. The functional integration over $\mathcal{A}_0$ gives the part of the state that is not seen by the detector. We can discard it, with the knowledge that the process is not trace preserving and the result needs to be normalized.

The functional integration over $\mathcal{M}_0$ produces the same result as the absurb case, giving a narrow function centered at the origin. If the Wigner function of the state $W[\alpha]$ does not overlap the origin, we can discard this part. However, many interesting states have Wigner functions sitting at the origin in phase space where they would be overlapped by this unwanted background term. In those cases, careful control of the modes that are detected can help to remove this unwanted term \cite{lvovfock}.

For the functional integration over $\mathcal{G}$, the integration is separated into an integration over the amplitude of $\Gamma$ and a functional integration over a field variable that is orthogonal to $\Gamma$. This separation is formally introduces with the aid of an {\em inhomogenous beamsplitter}. The transformation imposed by such an inhomogenous beamsplitter is represented by the substitutions
\begin{align}
\begin{split}
\alpha \rightarrow & P\diamond\alpha-i Q\diamond\beta \\
\beta \rightarrow & P\diamond\beta-i Q\diamond\alpha ,
\end{split}
\label{wigprojbstra}
\end{align}
where $P(\mathbf{k}_1,\mathbf{k}_2)=\Gamma(\mathbf{k}_1)\Gamma^*(\mathbf{k}_2)$ and $Q=\mathbf{1}-P$ are projection kernels. The transformation is performed on the Wigner functional of the state $W[\alpha]$, multiplied by that of a vacuum state, given by
\begin{equation}
W_{\text{vac}}[\beta]=\mathcal{N}_0 \exp(-2\|\beta\|^2) ,
\label{vacwig}
\end{equation}
where $\mathcal{N}_0$ is the normalization constant for a pure Gaussian state.
\begin{widetext}
When we apply Eq.~(\ref{wigprojbstra}) to Eq.~(\ref{wnaw}) after inserting a vacuum state, we obtain
\begin{align}
W_{\mathcal{G}}'(q,p) = & \mathcal{N} \int W'[\alpha,\beta]
\exp\left[ -\tfrac{1}{4} (\zeta^2+\xi^2) \Delta x^2\alpha^*\diamond P\diamond D\diamond P\diamond\alpha
-\tfrac{1}{4} (\zeta^2+\xi^2) \Delta x^2\beta^*\diamond Q\diamond D\diamond Q\diamond\beta \right. \nonumber \\
& \left. +\tfrac{1}{\sqrt{2}}(\zeta-i\xi)\Gamma^*\diamond D\diamond P\diamond\alpha
-\tfrac{1}{\sqrt{2}}(\zeta+i\xi)\alpha^*\diamond P\diamond D\diamond\Gamma +i q\xi-i p\zeta\right]\ \Dcirc[\alpha,\beta]\ \text{d}\zeta\ \text{d}\xi ,
\label{wnaww}
\end{align}
\end{widetext}
where
\begin{equation}
W'[\alpha,\beta] = W[P\diamond\alpha-i Q\diamond\beta] W_{\text{vac}}[P\diamond\beta-i Q\diamond\alpha] ,
\end{equation}
and we assumed that $\Gamma^*\diamond D\diamond Q=Q\diamond D\diamond\Gamma=0$.

The functional integral over $\alpha$ only contains a nontrivial state when the field variable is proportional to $\Gamma$. For the rest of the space, it is a vacuum state. The nontrivial part represents an ordinary integral over the complex valued amplitude of the field variable that is proportional to $\Gamma$. Hence, $P\diamond\alpha(\mathbf{k})\rightarrow \alpha_0 \Gamma(\mathbf{k})$, where $\alpha_0$ is a complex variable (not a {\em field} variable). The remaining part of the functional integration over $\alpha(\mathbf{k})$ produces a constant that is absorbed into the normalization constant $\mathcal{N}$.

The functional integral over $\beta$ can be separated in the same way. In this case, the state associated with the part of the field variable that is proportional to $\Gamma$ is a vacuum state. However, in this case, we retain the full space of the functional integral, because we need to maintain the invertibility of kernels that may appear in the Wigner functionals of the states.

When we apply these considerations, the expression in Eq.~(\ref{wnaww}) becomes
\begin{widetext}
\begin{align}
W_{\mathcal{G}}'(q,p) = & \mathcal{N} \int W[\beta](q_0,p_0) \exp\left[ -\tfrac{1}{8} \Delta x^2 \eta (\zeta^2+\xi^2) (q_0^2+p_0^2)
-\tfrac{1}{4} \Delta x^2 (\zeta^2+\xi^2) \beta^*\diamond D_{qq}\diamond\beta \right. \nonumber \\
& \left. +i q\xi-i p\zeta +i p_0\zeta\eta-i q_0\xi\eta \right]\ \Dcirc[\beta]\ \text{d}q_0\ \text{d}p_0\ \text{d}\zeta\ \text{d}\xi ,
\label{wnaw0}
\end{align}
\end{widetext}
where $\eta= \Gamma^*\diamond D\diamond\Gamma$ is the quantum efficiency of the detector, $D_{qq}=  Q\diamond D\diamond Q$, and we replaced the complex integration variable $\alpha_0$ with
\begin{equation}
\alpha_0 \rightarrow \tfrac{1}{\sqrt{2}}(q_0+i p_0) ,
\end{equation}
The functional integration therefore splits into a reduced functional integration that runs over the subspace $\mathcal{M}$ (i.e., those field variables that can pass through $D$) and an integration over the complex plane.

If we discard the $\Delta x^2$-terms in Eq.~(\ref{wnaw0}), we would get
\begin{align}
W_{\mathcal{G}}'(q,p) = & \mathcal{N} \int W[\beta](q_0,p_0) \exp\left[i (q-q_0\eta)\xi \right. \nonumber \\
& \left. -i (p-p_0\eta)\zeta \right]\ \Dcirc[\beta]\ \text{d}q_0\ \text{d}p_0\ \text{d}\zeta\ \text{d}\xi \nonumber \\
= & \mathcal{N} \int W[\beta]\left(\frac{q}{\eta},\frac{p}{\eta}\right)\ \Dcirc[\beta] .
\end{align}
The final functional integration over $\beta$ traces out all those degrees of freedom that are not associated with $\Gamma$. The result shows the effect of the detection efficiency $\eta$. It produces a scaling of the Wigner functional, which can be removed through a redefinition of the variables.

The separation of the different subspaces is governed by the nature of the detectors. There are different special cases that we can consider. Here, we'll consider two extreme cases: bucket detectors and single-mode detectors.

\subsection{Bucket detector}

If the detector is a bucket detector, then we can set $D(\mathbf{k}_1,\mathbf{k}_2)=\eta \mathbf{1}(\mathbf{k}_1,\mathbf{k}_2)$, where $\eta$ is the quantum efficiency of the detector, and $\mathbf{1}(\mathbf{k}_1,\mathbf{k}_2)$ is the identity.

In terms of the subspaces, we then have $\mathcal{M}_0\cong\mathcal{A}_0$, because all the elements in the functional phase space can be detected by the bucket detector. As a result, there are only two subspaces: $\mathcal{G}$ and $\mathcal{M}_0\cong\mathcal{A}_0$.

The effect on the expressions in Eq.~(\ref{absurd}) and Eq.~(\ref{wnaw0}) is that $\alpha^*\diamond D\diamond\alpha\rightarrow \eta\|\alpha\|^2$ and $\beta^*\diamond D_{qq}\diamond\beta\rightarrow\eta\beta^*\diamond Q\diamond\beta$, respectively. For further simplifications, we need to specify the initial Wigner functional. The coherent state is considered below as an example for this case.

\subsection{\label{enkelmodd}Single-mode detector kernel}

Alternatively, we consider $D$ as a single-mode detector kernel $D(\mathbf{k}_1,\mathbf{k}_2)=\eta M(\mathbf{k}_1)M^*(\mathbf{k}_2)$, where $M(\mathbf{k})$ is the normalized angular spectrum of the single mode. In this case, we'll assume that $M(\mathbf{k})=\Gamma(\mathbf{k})$. There are again only two subspaces: $\mathcal{A}_0$ and $\mathcal{G}\cong\mathcal{M}$. In this case, there is no equivalent for the absurd case in Eq.~(\ref{absurd}). The single-mode detector is preferred when the Wigner functional of the state overlaps the origin in phase space.

Since $\beta^*\diamond D_{qq}\diamond\beta=\beta^*\diamond Q\diamond\Gamma\Gamma^*\diamond Q\diamond\beta=0$, the integration over the subspace $\mathcal{G}$, with a subsequent normalization,
produces
\begin{align}
W_{\mathcal{G}}'(q,p) = & \frac{1}{(2\pi)^2} \int W_0(q_0,p_0) \exp\left[ i q\xi-i p\zeta \right. \nonumber \\
& -\tfrac{1}{8} \Delta x^2 \eta (\zeta^2+\xi^2) (q_0^2+p_0^2) \nonumber \\
& \left. +i p_0\zeta\eta-i q_0\xi\eta \right]\ \text{d}q_0\ \text{d}p_0\ \text{d}\zeta\ \text{d}\xi ,
\label{wnaw1}
\end{align}
where we traced over $\beta$, and defined
\begin{equation}
\int W[\beta](q_0,p_0)\ \Dcirc[\beta] = W_0(q_0,p_0) .
\end{equation}
The integrations over $\zeta$ and $\xi$ evaluate to
\begin{align}
W_{\mathcal{G}}'(q,p) = & \int
\exp\left[-2\frac{(q_0\eta-q)^2+(p_0\eta-p)^2}{(q_0^2+p_0^2)\Delta x^2\eta}\right] \nonumber \\
& \times \frac{2 W_0(q_0,p_0)}{(q_0^2+p_0^2)\pi\Delta x^2\eta}\ \text{d}q_0\ \text{d}p_0 .
\label{wnawkern}
\end{align}
The observed Wigner function is thus obtained from the traced Wigner functional through a linear integral operation (superposition integral) with a kernel given by
\begin{align}
\kappa(q_0,p_0,q,p) = &
\exp\left[-2\frac{(q_0\eta-q)^2+(p_0\eta-p)^2}{(q_0^2+p_0^2)\Delta x^2\eta}\right] \nonumber \\
& \times \frac{2}{(q_0^2+p_0^2)\pi\Delta x^2\eta} .
\label{homokern}
\end{align}

There is also a scaling introduced by the quantum efficiency $\eta$. This scaling can be removed from Eq.~(\ref{wnawkern}) through the redefinitions $\{q,p\}\rightarrow\{q'\eta,p'\eta\}$, and a renormalization, leading to
\begin{align}
W_{\mathcal{G}}'(q',p') = & \int
\exp\left[-2\eta\frac{(q_0-q')^2+(p_0-p')^2}{(q_0^2+p_0^2)\Delta x^2}\right] \nonumber \\
& \times \frac{2\eta W_0(q_0,p_0)}{(q_0^2+p_0^2)\pi\Delta x^2}\ \text{d}q_0\ \text{d}p_0 .
\label{wnaws}
\end{align}
The quantum efficiency is now associated with $\Delta x$, and represents a slight reduction in the effective number of photons in the local oscillator.

Without the factors of $q_0^2+p_0^2$ in the denominators, Eq.~(\ref{homokern}) would represent a Dirac delta function in the limit $\Delta x\rightarrow 0$. However, the factors of $q_0^2+p_0^2$ in the denominators make the kernel dependent on the distance from the origin. When $q=p=0$, the kernel is severely singular at the origin as a function of $\{q_0,p_0\}$. For fixed values of $\{q,p\}>0$, and a small value for $\Delta x$, the kernel gives a narrow Gaussian peak located at $\{q_0,p_0\}=\{q\eta^{-1},p\eta^{-1}\}$. It becomes broader as the point $\{q,p\}$ moves further away from the origin.

In fact, the kernel has a scale invariance: we can multiply all the variables by the same factor and it will cancel apart from an overall change in the amplitude of the kernel. It implies that the width of the peak scales linearly with the distance of the peak from the origin. The peak would thus become comparable to the minimum uncertainty area when $q_0^2+p_0^2\sim\zeta_0^2$ --- i.e., when the average number of photons in the state becomes comparable to the average number of photons in the local oscillator.

Due to the factor of $q_0^2+p_0^2$ in the denominators, the integrals in Eq.~(\ref{wnawkern}) tend to be intractable. If $\Delta x$ is small enough, we can argue that for $\{q,p\}>0$, the kernel becomes zero whenever $\{q_0,p_0\}$ differs by more than $\Delta x$ from the location of its peak. Therefore, we can substitute $q_0^2+p_0^2\rightarrow (q^2+p^2)\eta^{-2}$, which makes the integration over $\{q_0,p_0\}$ more tractable.

The expression in Eq.~(\ref{wnawkern}) then becomes
\begin{align}
W_{\mathcal{G}}'(q,p) \approx & \int
\exp\left[-2\eta\frac{(q_0\eta-q)^2+(p_0\eta-p)^2}{(q^2+p^2)\Delta x^2}\right] \nonumber \\
& \times \frac{2\eta W_0(q_0,p_0)}{(q^2+p^2)\pi\Delta x^2}\ \text{d}q_0\ \text{d}p_0 ,
\label{wnawkern0}
\end{align}
which is now similar to a convolusion, where the resolution of the observed Wigner function is determined by the ratio of the average number of photons in the state to the average number of photons in the local oscillator after the reduction imposed by the detection efficiency.

\section{\label{kohfgs}Example: coherent state}

As a first example, we consider the homodyne tomography of an arbitrary coherent state. The transformation of the inhomogenous beamsplitter is performed on the Wigner functional state times that of a vacuum state by substituting Eq.~(\ref{wigprojbstra}) into the combined Wigner functional of the state and the vacuum. The effect is
\begin{align}
W_{\text{coh}}[\alpha,\beta] = & \mathcal{N}_0^2 \exp\left(-2\|\alpha-\varphi\|^2-2\|\beta\|^2\right) \nonumber \\
\rightarrow & \mathcal{N}_0\exp\left(-2\|\alpha-P\diamond\varphi\|^2\right) \nonumber \\
& \times \mathcal{N}_0\exp\left(-2\|\beta-Q\diamond\varphi\|^2\right) ,
\end{align}
where $\varphi(\mathbf{k})$ is the spectral parameter function of the coherent state. After we trace out the degrees of freedom of $\alpha$ that are orthogonal to $\Gamma$, the result reads
\begin{align}
W_{\text{coh}}[\beta](\alpha_0) = & 2\exp\left(-2|\alpha_0-\alpha_1|^2\right) \nonumber \\
& \times \mathcal{N}_0\exp\left(-2\|\beta-\beta_1\|^2\right) ,
\label{sepkoh}
\end{align}
where $\alpha_1= \Gamma^*\diamond\varphi$ is the complex coefficient for the part of $\varphi$ proportional to $\Gamma$, and $\beta_1=  Q\diamond\varphi$ is a complex function representing the part of $\varphi$ that is orthogonal to $\Gamma$. After substituting Eq.~(\ref{sepkoh}) into Eq.~(\ref{wnaw0}), we obtain
\begin{widetext}
\begin{align}
W_{\text{coh}}(q,p) = & \frac{\mathcal{N}_0}{2\pi^2} \int \exp\left[-(q_0-q_1)^2-(p_0-p_1)^2
-\tfrac{1}{8} \Delta x^2 \eta (\zeta^2+\xi^2) (q_0^2+p_0^2) +i q\xi-i p\zeta +i p_0\zeta\eta-i q_0\xi\eta\right] \nonumber \\
& \times \exp\left[-2\|\beta-\beta_1\|^2-\tfrac{1}{4} \Delta x^2 (\zeta^2+\xi^2) \beta^*\diamond D_{qq}\diamond\beta\right]\ \Dcirc[\beta]\ \text{d}q_0\ \text{d}p_0\ \text{d}\zeta\ \text{d}\xi ,
\label{wnawkohi}
\end{align}
\end{widetext}
where we expressed $\alpha_0$ in terms of $q_0$ and $p_0$, and replaced $\alpha_1 \rightarrow \tfrac{1}{\sqrt{2}}(q_1+i p_1)$. The integrations over $q_0$ and $p_0$ are separated from the functional integration over $\beta$.

\subsection{Bucket detector}

For the bucket detector, we replace $D_{qq}=Q\diamond D\diamond Q\rightarrow \eta Q$ in Eq.~(\ref{wnawkohi}). We evaluate the functional integration over $\beta$ and also perform the integrations over $q_0$ and $p_0$, to obtain
\begin{align}
W_{\mathcal{G}}'(q,p) = & \int \exp\left[-\tfrac{1}{4} \frac{(\xi\eta+i 2 q_1)^2+(\zeta\eta-i 2 p_1)^2}{1+\tau} \right. \nonumber \\
& \left. -q_1^2-p_1^2+i q\xi-i p\zeta -2\frac{\tau}{1+\tau} \|\beta_1\|^2 \right] \nonumber \\
& \times \frac{1}{2\pi(1+\tau)^{\Omega}}\ \text{d}\zeta\ \text{d}\xi ,
\end{align}
where $\Omega=\tr\{Q\}+1$, and
\begin{equation}
\tau = \tfrac{1}{8} (\zeta^2+\xi^2)\eta \Delta x^2 .
\end{equation}

Since $\tau$ contains the radial dependence of the remaining integration variables, the factor of $1/(1+\tau)^{\Omega}$ restricts the integration domain that would contribute to a region close to the origin. Therefore, we can set $1+\tau\rightarrow 1$, and evaluate the remaining integration. Hence,
\begin{align}
W_{\mathcal{G}}'(q,p) \approx & \frac{1}{2\pi} \int \exp\left[-\tfrac{1}{4} (\xi\eta+i 2 q_1)^2-\tfrac{1}{4}(\zeta\eta-i 2 p_1)^2 \right. \nonumber \\
& -q_1^2-p_1^2+i q\xi-i p\zeta  \nonumber \\
& \left. -\tfrac{1}{4} (\zeta^2+\xi^2)\eta \Delta x^2 \|\beta_1\|^2 \right]\ \text{d}\zeta\ \text{d}\xi \nonumber \\
= & \frac{2}{\eta^2+\eta \Delta x^2 \|\beta_1\|^2} \nonumber \\
& \times \exp\left[-2\frac{|\alpha-\eta\alpha_1|^2}{\eta^2+\eta \Delta x^2 \|\beta_1\|^2} \right] ,
\end{align}
where we expressed the result in terms of $\alpha$'s at the end. If we set $\Delta x=0$, the result is a scaled version of the original coherent state. We can compensate for the scaling by redefining the variable $\alpha\rightarrow\alpha'\eta$ and renormalizing the function. The result becomes
\begin{equation}
W_{\mathcal{G}}'(\alpha') = \frac{2}{1+\Delta w} \exp\left(\frac{-2|\alpha'-\alpha_1|^2}{1+\Delta w} \right) ,
\label{kohemmer}
\end{equation}
where
\begin{equation}
\Delta w =  \frac{\Delta x^2 \|\beta_1\|^2}{\eta} = \frac{\|Q\diamond\varphi\|^2}{\eta\zeta_0^2} .
\end{equation}
We see that the width of the rescaled state is increased by the ratio of the number of photons that can pass through $Q$ over the number of photons in the local oscillator, reduced by the quantum efficiency.

\subsection{Single-mode detector kernel}

For a single-mode detector with $M(\mathbf{k})=\Gamma(\mathbf{k})$, we get $\beta^*\diamond D_{qq}\diamond\beta=|\Gamma^*\diamond Q\diamond\beta|^2=0$. The functional integration over $\beta$ can be evaluated without complications. So, Eq.~(\ref{wnawkohi}) becomes
\begin{align}
W_{\text{coh}}(q,p) = & \frac{1}{2\pi^2} \int \exp\left[-(q_0-q_1)^2-(p_0-p_1)^2 \right. \nonumber \\
& \left. -\tfrac{1}{8} \Delta x^2 \eta (\zeta^2+\xi^2) (q_0^2+p_0^2) +i q\xi-i p\zeta \right. \nonumber \\
& \left. +i p_0\zeta\eta-i q_0\xi\eta \right]\ \text{d}q_0\ \text{d}p_0\ \text{d}\zeta\ \text{d}\xi  .
\end{align}
If we first evaluate the integration over $q_0$ and $p_0$, as with the bucket detector case, we'll again get factors of $1+\tau$ in the denominated, but this time the dependence is not as severely suppressed, which implies that the approximation $1+\tau\approx 1$ is not as valid. Therefore, we first integrate over $\zeta$ and $\xi$ to obtain
\begin{align}
W_{\mathcal{G}}'(q,p) = & \int 4\exp\left[-(q_0-q_1)^2-(p_0-p_1)^2\right] \nonumber \\
& \times \exp\left[-2\frac{(q_0\eta-q)^2+(p_0\eta-p)^2}{(q_0^2+p_0^2)\eta\Delta x^2}\right] \nonumber \\
& \times \frac{1}{(q_0^2+p_0^2)\pi\eta\Delta x^2}\ \text{d}q_0\ \text{d}p_0 ,
\end{align}
which corresponds to Eq.~(\ref{wnawkern}). It can be assumed that the kernel peak is narrow enough for small $\Delta x$ so that we can substitute $q_0^2+p_0^2\rightarrow (q^2+p^2)\eta^{-2}$, as before. The integrals over $q_0$ and $p_0$ can then be evaluated to give
\begin{align}
W_{\mathcal{G}}'(q,p) = & \frac{2\eta}{\eta^3+|\alpha|^2\Delta x^2} \nonumber \\
& \times \exp\left(-2\frac{\eta|\alpha-\eta\alpha_1|^2}{\eta^3+|\alpha|^2\Delta x^2}\right) ,
\end{align}
where we converted the expression back to complex valued variables. We recover a scaled version of the Wigner function for the coherent states, but with a different width. If we set $\Delta x=0$, the result is a scaled version of the original coherent state due to the reduced efficiency represented by $\eta$. Compensating for the scaling by redefining the complex variable $\alpha\rightarrow\alpha'\eta$, we obtain
\begin{align}
W_{\mathcal{G}}'(\alpha') = & \frac{2}{1+\frac{1}{\eta}|\alpha'|^2\Delta x^2} \nonumber \\
& \times \exp\left(-2\frac{|\alpha'-\alpha_1|^2}{1+\frac{1}{\eta}|\alpha'|^2\Delta x^2}\right) .
\end{align}
For large enough $|\alpha_1|$, we can replace $|\alpha'|^2\rightarrow|\alpha_1|^2$ in the denominators. The result then has the same form as in Eq.~(\ref{kohemmer}), but this time, the increase in width is given by the ratio of the average number of photons in the state that can be observed by the detector to the reduced average number of photons in the local oscillator:
\begin{equation}
\Delta w = \frac{\Delta x^2 |\alpha_1|^2}{\eta} = \frac{|\Gamma^*\diamond\varphi|^2}{\eta\zeta_0^2} .
\end{equation}

\section{Example: Fock states}

Since the Wigner functionals of Fock states are centered at the origin of phase space, we only consider the single-mode detector. The generating function for the Wigner functionals of the single-mode Fock states is
\begin{equation}
\mathcal{W} = \frac{\mathcal{N}_0}{1+J} \exp\left(-2\|\alpha\|^2+\frac{4J}{1+J}
\alpha^*\diamond FF^*\diamond\alpha\right) ,
\label{genfock}
\end{equation}
where $F(\mathbf{k})$ represents the normalized angular spectral parameter function for the Fock states, and $J$ is the generating parameter. After combining it with the Wigner functional for the vacuum state in Eq.~(\ref{vacwig}), and applying Eq.~(\ref{wigprojbstra}) to separate the integration domains, we obtain
\begin{align}
\mathcal{W}[\alpha,\beta](J)
 = & \frac{\mathcal{N}_0^2}{1+J} \exp\left[i 2\mathcal{H}\beta^*\diamond Q\diamond FF^*\diamond P\diamond\alpha \right. \nonumber \\
& -i 2\mathcal{H}\alpha^*\diamond P\diamond FF^*\diamond Q\diamond\beta \nonumber \\
& -2\alpha^*\diamond \left(\mathbf{1}-\mathcal{H}P\diamond FF^*\diamond P\right)\diamond\alpha \nonumber \\
& \left. -2\beta^*\diamond \left(\mathbf{1}-\mathcal{H}Q\diamond FF^*\diamond Q\right)\diamond\beta \right] ,
\end{align}
where
\begin{equation}
\mathcal{H} = \frac{2J}{1+J} .
\end{equation}

Here, we are interested in the case when the parameter function of the Fock states does not exactly match the mode of the local oscillator. Therefore, we assume that $F(\mathbf{k}) = \mu\Gamma(\mathbf{k}) + \nu\Lambda(\mathbf{k})$, where $|\mu|^2+|\nu|^2=1$, $\Gamma^*\diamond\Lambda=P\diamond\Lambda=0$ and $Q\diamond\Lambda=\Lambda$. As a result, $|\mu|^2$ is the overlap efficiency. After integrating out the part of the $\alpha$-dependent functional orthogonal to $\Gamma$, we obtain
\begin{align}
\mathcal{W}[\beta](\alpha_0,J) = & \frac{2\mathcal{N}_0}{1+J}
\exp\left[-2\left(1-\mathcal{H}|\mu|^2\right)|\alpha_0|^2 \right. \nonumber \\
& + i 2\mathcal{H}\mu^*\nu\alpha_0 \beta^*\diamond\Lambda -i 2\mathcal{H}\mu\nu^*\alpha_0^* \Lambda^*\diamond\beta \nonumber \\
& \left. -2\beta^*\diamond\mathcal{K}\diamond\beta \right] .
\end{align}
where
\begin{equation}
\mathcal{K} = \mathbf{1}-\mathcal{H}|\nu|^2 \Lambda\Lambda^* .
\end{equation}

The functional integration over $\beta$ implies tracing the state over $\beta$, which produces
\begin{align}
\mathcal{W}(\alpha_0,J) = & \frac{2}{(1+J)\det\{\mathcal{K}\}}
 \exp\left[-2\left(1-\mathcal{H}|\mu|^2\right)|\alpha_0|^2 \right. \nonumber \\
 & \left. +2 \mathcal{H}^2 |\mu|^2|\nu|^2|\alpha_0|^2 \Lambda^*\diamond\mathcal{K}^{-1}\diamond\Lambda \right] .
\end{align}
The determinant and inverse can be simplified as
\begin{align}
\begin{split}
\det\left\{\mathcal{K}\right\}=\det\left\{\mathbf{1}-\mathcal{H}|\nu|^2 \Lambda\Lambda^*\right\} = & 1-\mathcal{H}|\nu|^2 , \\
\mathcal{K}^{-1}=\left(\mathbf{1}-\mathcal{H}|\nu|^2 \Lambda\Lambda^*\right)^{-1}
= & \mathbf{1}+\frac{\mathcal{H}|\nu|^2\Lambda\Lambda^*}{1-\mathcal{H}|\nu|^2} .
\end{split}
\end{align}
Therefore, the expression becomes
\begin{equation}
\mathcal{W}(\alpha_0,J) = \frac{2\exp\left(-2|\alpha_0|^2\right)}{1+J\omega}
\exp\left(\frac{4J|\mu|^2|\alpha_0|^2}{1+J\omega}\right) ,
\label{genfock0}
\end{equation}
where we used $|\nu|^2=1-|\mu|^2$ to define
\begin{equation}
\omega =  1-2|\nu|^2=2|\mu|^2-1=|\mu|^2-|\nu|^2 .
\end{equation}

We replace $W_0(q_0,p_0)$ in Eq.~(\ref{wnawkern0}) by the generating function in Eq.~(\ref{genfock0}) to compute a generating function for the observed Wigner functions of the Fock states:
\begin{align}
\mathcal{W}_F(\alpha,J) = & \int \exp\left[-\frac{1+J\omega-2J|\mu|^2}{1+J\omega}(q_0^2+p_0^2)\right] \nonumber \\
& \times \exp\left[-2\eta\frac{(q_0\eta-q)^2+(p_0\eta-p)^2}{(q^2+p^2)\Delta x^2}\right] \nonumber \\
& \times \frac{4\eta}{(1+J\omega)(q^2+p^2)\pi\Delta x^2}\ \text{d}q_0\ \text{d}p_0 \nonumber \\
 = & \exp\left[-\frac{2(1-J)|\alpha|^2\eta}{(1-J)|\alpha|^2\Delta x^2+(1+J\omega)\eta^3}\right] \nonumber \\
& \times \frac{2\eta}{(1-J)|\alpha|^2\Delta x^2+(1+J\omega)\eta^3} .
\end{align}
The expression already incorporates the approximation where we set $q_0^2+p_0^2\rightarrow (q^2+p^2)\eta^{-2}$ in the denominator.

\begin{figure}[ht]
\centerline{\includegraphics{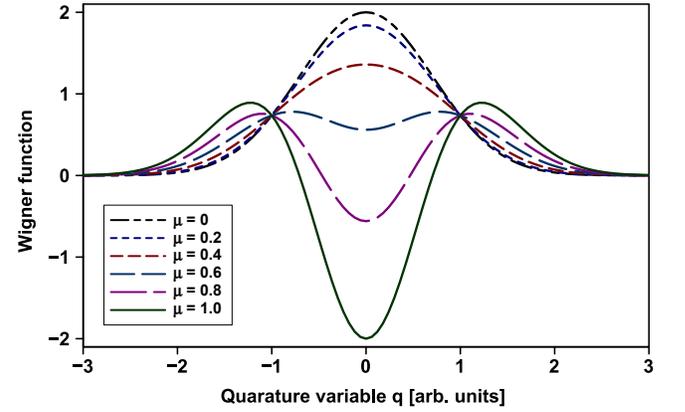}}
\caption{Observed Wigner function of a single-photon Fock state as a function $q$ with $p=0$ for different values of $|\mu|$.}
\label{fockwig}
\end{figure}

Since the Wigner functions of the Fock states are located at the origin, we can assume that $|\alpha|^2\Delta x^2\ll 1$. Therefore, we can set $\Delta x^2=0$. The expression then simplifies to
\begin{align}
\mathcal{W}(\alpha,J) = & \frac{2}{(1+J\omega)\eta^2} \nonumber \\
& \times \exp\left[-\frac{2|\alpha|^2}{\eta^2}+\frac{4J|\mu|^2|\alpha|^2}{(1+J\omega)\eta^2}\right] .
\label{fockwiggen}
\end{align}
When we redefine $\alpha\rightarrow\alpha'\eta$ to remove that scaling caused by $\eta$, we recover Eq.~(\ref{genfock0}). The Wigner functions of the individual Fock states are then given by
\begin{align}
W_{\ket{n}\bra{n}}(\alpha') = & (1-2|\mu|^2)^n \exp\left(-2|\alpha'|^2\right) \nonumber \\
& \times L_n\left(\frac{4|\mu|^2|\alpha'|^2}{2|\mu|^2-1}\right) ,
\label{focknwig}
\end{align}
where $L_n(\cdot)$ is the $n$-th order Laguerre polynomial. For $|\mu|<1$, it is scaled relative to the Gaussian envelope. In Fig.~\ref{fockwig}, we show the observed Wigner function of a single-photon Fock state for different values of $|\mu|$, ranging from that of the Fock state (for $|\mu|=1$) to that of a vacuum state (for $|\mu|=0$).

\subsection{Marginal distributions}

We can use the generating function in Eq.~(\ref{genfock0}) to investigate the marginal distributions of the Wigner function that it produces. The variable $\alpha$ is expressed in terms of $q$ and $p$, and the resulting expression is integrated over $p$ to produce a generating function for the observed marginal distributions, which is given by
\begin{align}
\mathcal{W}(q,J) = & \int \mathcal{W}(q,p,J)\ \frac{\text{d}p}{2\pi} \nonumber \\
 = & \frac{1}{\sqrt{\pi(1-J)(1+J\omega)}} \nonumber \\
& \times \exp\left(-q^2+\frac{2J|\mu|^2q^2}{1+J\omega}\right) .
\label{margnwig}
\end{align}
The observed marginal distribution for the single-photon Fock state is
\begin{equation}
\left. \partial_J \mathcal{W}(q,J) \right|_{J=0} = \frac{\exp(-q^2)}{\sqrt{\pi}} \left(2|\mu|^2q^2+1-|\mu|^2\right) .
\end{equation}
It is a non-negative function for all the allowed values of $|\mu|$ (i.e., $0\leq |\mu|\leq 1$). For $|\mu|=1$, the distribution is zero at the origin, but for smaller values of $|\mu|$ it is larger than zero at the origin.

We can compare this result with what would be obtained from a naive approach where we simply substitute $\alpha(\mathbf{k})\rightarrow\alpha \Gamma(\mathbf{k})$ into the generating function for Wigner functionals of the Fock states, to get
\begin{equation}
\mathcal{W}'(\alpha,J) = \frac{2}{1+J}\exp\left(-2|\alpha|^2+\frac{4J|\mu|^2}{1+J}|\alpha|^2\right) ,
\end{equation}
instead of Eq.~(\ref{genfock0}). After applying the same integration over $p$ to produce the generating function for the marginal distributions, we obtain
\begin{align}
\mathcal{W}'(q,J) = &  \frac{1}{\sqrt{\pi(1+J)(1-J\omega)}} \nonumber \\
& \times \exp\left(-q^2+\frac{2J|\mu|^2q^2}{1+J}\right) .
\end{align}
In this case, the marginal distribution for the single-photon Fock state is
\begin{equation}
\left. \partial_J \mathcal{W}'(q,J) \right|_{J=0} = \frac{\exp(-q^2)}{\sqrt{\pi}} \left(2|\mu|^2q^2-1+|\mu|^2\right) .
\end{equation}
At the origin, this function is negative for $|\mu|^2<1$, which represents a non-physical situation. Therefore, the naive approach does not in general give valid Wigner functions.

\section{\label{www}Example: squeezed vacuum state}

As a final example, we consider the homodyne tomography process of a squeezed vacuum state, using single-mode detection. A pure squeezed vacuum state has a Wigner functional given by
\begin{align}
W_{\text{sv}}[\alpha] = & \mathcal{N}_0 \exp\left(-2\alpha^*\diamond A\diamond\alpha \right. \nonumber \\
& \left. -\alpha^*\diamond B\diamond\alpha^*-\alpha\diamond B^*\diamond\alpha\right) ,
\label{squ}
\end{align}
where $A$ and $B$ are kernel functions depending on a squeezing parameter $\Xi$.

If we naively express the observed Wigner function as that which is obtained by subtituting $\alpha(\mathbf{k})\rightarrow \alpha_0\Gamma(\mathbf{k})$ into Eq.~(\ref{squ}), it would read
\begin{equation}
W_{\text{nsv}}(\alpha_0) = \mathcal{N}\exp\left(-2|\alpha_0|^2 g_A-\alpha_0^{*2} g_B-\alpha_0^2 g_B^*\right) ,
\label{nsqu}
\end{equation}
where $\mathcal{N}$ is a normalization constant, and
\begin{align}
\begin{split}
g_A =  & \Gamma^*\diamond A\diamond\Gamma , \\
g_B =  & \Gamma^*\diamond B\diamond\Gamma^* , \\
g_B^* =  & \Gamma\diamond B^*\diamond\Gamma .
\end{split}
\label{koefqq}
\end{align}
However, we will see below that the trace over $\beta$ can introduce distortions to this function.

We first perform the separation of the functional phase space by using the transformation given in Eq.~(\ref{wigprojbstra}). Then, we integrate out the part of the state that depends on $\alpha$ and is orthogonal to $\Gamma$. The result is
\begin{align}
W_{\text{sv}}[\beta](\alpha_0) = & 2\exp\left(-2|\alpha_0|^2 g_A-\alpha_0^{*2} g_B-\alpha_0^2 g_B^*\right) \nonumber \\
 & \times \mathcal{N}_0 \exp\left[ -2\beta^*\diamond A_q\diamond\beta +\beta^*\diamond B_{qq}\diamond\beta^* \right. \nonumber \\
& +\beta\diamond B_{qq}^*\diamond\beta -i 2\beta^*\diamond (U\alpha_0+V\alpha_0^*) \nonumber \\
& \left. +i 2(\alpha_0^* U^*+\alpha_0 V^*)\diamond\beta \right] ,
 \label{squ2}
\end{align}
where
\begin{align}
\begin{split}
E_{qq} =  & Q\diamond E\diamond Q, \\
B_{qq} =  & Q\diamond B\diamond Q^* , \\
B_{qq}^* =  & Q^*\diamond B^*\diamond Q , \\
A_q =  & \mathbf{1}+E_{qq} ,
\end{split}
\label{subqq}
\end{align}
with $E=A-\mathbf{1}$, and
\begin{align}
\begin{split}
U =  & Q\diamond E\diamond\Gamma , \\
V =  & Q\diamond B\diamond\Gamma^* ,
\end{split}
\end{align}
are functions orthogonal to $\Gamma$. They are included because $\Gamma$ is generally not an eigenfunction of the kernels. The kernels transform $\Gamma$ as follows:
\begin{align}
\begin{split}
E\diamond\Gamma = & P\diamond E\diamond\Gamma + Q\diamond E\diamond\Gamma = g_E\Gamma + U , \\
B\diamond\Gamma^* = & P\diamond B\diamond\Gamma^* + Q\diamond B\diamond\Gamma^* = g_B\Gamma + V ,
\end{split}
\label{subgam}
\end{align}
where $g_E =  \Gamma^*\diamond E\diamond\Gamma = g_A-1$.

The first line in Eq.~(\ref{squ2}) contains the result that we obtained from the naive approach, given in Eq.~(\ref{nsqu}). Hence, we can represent Eq.~(\ref{squ2}) as
\begin{equation}
W_{\text{sv}}[\beta](\alpha_0) = W_{\text{nsv}}(\alpha_0) W_{\beta}[\beta](\alpha_0) .
\end{equation}

The single-mode detector with $M(\mathbf{k})=\Gamma(\mathbf{k})$ leads to $\beta^*\diamond D_{qq}\diamond\beta=0$. Therefore, the functional integral over $\beta$ implies the trace of the state over $\beta$. Considering only the $\beta$-dependent part of the expression, we obtain
\begin{align}
W_{\beta}'(\alpha_0) = & \int W_{\beta}[\beta](\alpha_0)\ \Dcirc[\beta] \nonumber \\
 = & \left(\det\{A_q\}\det\{K\}\right)^{-1/2} \exp\left[\psi^*\diamond A_q^{-1}\diamond\psi \right. \nonumber \\
& + \left(\psi-\psi^*\diamond A_q^{-1}\diamond B_{qq}\right) \diamond K^{-1} \nonumber \\
& \left. \diamond \left(\psi^*-B_{qq}^*\diamond A_q^{-1}\diamond \psi\right) \right] ,
\label{squb}
\end{align}
where
\begin{align}
\begin{split}
\psi = & U\alpha_0+V\alpha_0^* , \\
K = & A_q^*-B_{qq}^*\diamond A_q^{-1}\diamond B_{qq} .
\end{split}
\end{align}

The result in Eq.~(\ref{squb}) can be represented as
\begin{equation}
W_{\beta}'(\alpha_0) = \mathcal{N}_{\beta} \exp\left(2|\alpha_0|^2 h_A+\alpha_0^{*2} h_B+\alpha_0^2 h_B^*\right) ,
\label{squb0}
\end{equation}
where
\begin{align}
\begin{split}
\mathcal{N}_{\beta} =  & \left(\det\{A_q\} \det\{K\}\right)^{-1/2} , \\
h_A =  & U^*\diamond A_q^{-1}\diamond U + V^*\diamond A_q^{-1}\diamond V \\
& + \Psi_u^*\diamond K^{-1}\diamond\Psi_u + \Psi_v^*\diamond K^{-1} \diamond\Psi_v , \\
h_B =  & U^*\diamond A_q^{-1}\diamond V + \Psi_u^*\diamond K^{-1}\diamond\Psi_v , \\
h_B^* =  & V^*\diamond A_q^{-1}\diamond U + \Psi_v^*\diamond K^{-1}\diamond \Psi_u .
\end{split}
\label{distor}
\end{align}
with
\begin{align}
\begin{split}
\Psi_u =  & B_{qq}^*\diamond A_q^{-1}\diamond U - V^* , \\
\Psi_v =  & B_{qq}^*\diamond A_q^{-1}\diamond V - U^* .
\end{split}
\label{psidefs}
\end{align}

The combination of Eq.~(\ref{squb0}) with the $\beta$-independent part of Eq.~(\ref{squ2}) becomes
\begin{align}
W_{\text{sv}}'(\alpha_0) = & 2\mathcal{N}_{\beta} \exp\left[-2|\alpha_0|^2 (g_A-h_A) \right. \nonumber \\
& \left. -\alpha_0^{*2} (g_B-h_B)-\alpha_0^2 (g_B^*-h_B^*)\right] .
\label{squ3}
\end{align}
Since the $\beta$-dependent part of Eq.~(\ref{squ2}) also contains $\alpha_0$, the trace over $\beta$ generally produces an $\alpha_0$-dependent function that modifies $W_{\text{nsv}}(\alpha_0)$ and thereby distorts it.

The observed Wigner function is determined by substituting Eq.~(\ref{squ3}) in the place of $W_0(q_0,p_0)$ in Eq.~(\ref{wnawkern0}), which assumes a small $\Delta x^2$. Here, we set $\eta=1$, because the effect of $\eta$ is the same as in the previous cases. After evaluating the integrals, we obtain
\begin{align}
W_{\text{osv}}(\alpha)
 = & \frac{2\mathcal{N}_{\beta}}{\sqrt{g_D}} \exp\left(-2|\alpha|^2 g_C -2|\alpha|^2 \frac{g_A-h_A-g_C}{g_D} \right. \nonumber \\
& \left. -\alpha^{*2} \frac{g_B-h_B}{g_D}-\alpha^2 \frac{g_B^*-h_B^*}{g_D} \right) ,
\end{align}
where we discarded the $\Delta x^4$-terms, and defined
\begin{align}
\begin{split}
g_C = & \frac{(g_A-h_A)^2-|g_B-h_B|^2}{2(g_A-h_A)} , \\
g_D = & 1+2|\alpha|^2(g_A-h_A)\Delta x^2 .
\end{split}
\end{align}

If we set $\Delta x=0$, the expression becomes the same as in Eq.~(\ref{squ3}). Therefore, the distortions would not be removed by increasing the power in the local oscillator.

\subsection{Weakly squeezed vacuum state}

The complexity of the expression in Eq.~(\ref{squ3}), as represented by the quantities in Eq.~(\ref{distor}), indicates that the observed Wigner function of a squeezed vacuum state could in general be severely distorted. However, it may be reasonable to expect that the distortions would be reduced if the state is only weakly squeezed. To investigate this possibility, we'll consider a squeezing parameter $\Xi$ that is small. Then we can expand the kernels and keep only terms up to second order in $\Xi$. As a result, $A \approx \mathbf{1}+E_2$ and $A^{-1} \approx \mathbf{1}-E_2$, where $E_2$ is second order in $\Xi$. Moreover, $B$ and $B^*$ are first order in $\Xi$. We also define $U=g_U U_0$ and $V=g_V V_0$, so that
\begin{align}
\begin{split}
E_2\diamond\Gamma = & g_E\Gamma + g_U U_0 , \\
B\diamond\Gamma^* = & g_B\Gamma + g_V V_0 ,
\end{split}
\end{align}
where $U_0$ and $V_0$ are normalized functions.

By replacing $Q\rightarrow\mathbf{1}+\Gamma\Gamma^*$, and using Eqs.~(\ref{subqq}) and (\ref{subgam}), we have
\begin{align}
\begin{split}
A_q = & \mathbf{1} + E_2 - g_E\Gamma\Gamma^* - g_U U_0\Gamma^* - g_U^*\Gamma U_0^* , \\
A_q^{-1} \approx & \mathbf{1} - E_2 + g_E\Gamma\Gamma^* + g_U U_0\Gamma^* + g_U^*\Gamma U_0^* .
\end{split}
\end{align}
The purity of the initial squeezed vacuum states implies that, to second order in $\Xi$,
\begin{equation}
B\diamond B^* \approx 2E_2 .
\label{ebb}
\end{equation}
Therefore, the expressions for $K$ and its inverse become
\begin{align}
\begin{split}
K \approx & \mathbf{1} - E_2 + g_E\Gamma^*\Gamma \\
& + g_U \Gamma^* U_0 + g_U^* U_0^*\Gamma + |g_V|^2 V_0^* V_0 , \\
K^{-1} \approx & \mathbf{1} + E_2 - g_E\Gamma^*\Gamma  \\
& - g_U \Gamma^* U_0 - g_U^* U_0^*\Gamma - |g_V|^2 V_0^* V_0 .
\end{split}
\end{align}
To second order in $\Xi$, the product of determinants is
\begin{align}
\det\{A_q\} \det\{K\} = & \det\{A_q\diamond K\} \nonumber \\
\approx & \det\left\{\mathbf{1}+|g_V|^2 V_0 V_0^*\right\} \nonumber \\
 = & 1 + |g_V|^2 .
\end{align}
Here, it is assumed that $|g_V|<1$, otherwise the expansion would not be convergent. Although the identity $\mathbf{1}$ is infinite dimensional, by itself it just gives $1^{\Omega}=1$. The only part that deviates from $\mathbf{1}$ is one-dimensional. Therefore, the power becomes 1.

Since the leading contribution in $\psi$ is first order in $\Xi$, the expansion of the exponent in Eq.~(\ref{squb}) to second order in $\Xi$ implies that the inverses become $A_q^{-1}\rightarrow\mathbf{1}$ and $K^{-1}\rightarrow\mathbf{1}$. Moreover, all the terms in Eq.~(\ref{distor}) that contain $U$'s are dropped, because they are already second order in $\Xi$.

The first term in the exponent in Eq.~(\ref{squb}) becomes
\begin{equation}
\psi^*\diamond A_q^{-1}\diamond\psi \approx \psi^*\diamond\psi \approx |\alpha_0|^2 |g_V|^2 ,
\end{equation}
to second order in $\Xi$. Since $\psi$ and $B_{qq}$ are first order in $\Xi$ and orthogonal to $\Gamma$, it follows that
\begin{align}
\begin{split}
B_{qq}^*\diamond A_q^{-1}\diamond\psi \approx & g_V B^*\diamond V_0 \alpha_0^* , \\
\psi^*\diamond A_q^{-1}\diamond B_{qq} \approx & g_V^* V_0^*\diamond B \alpha ,
\end{split}
\end{align}
which are at least second order in $\Xi$. Therefore, the second term in the exponent also becomes
\begin{align}
\left(\psi-\psi^*\diamond A_q^{-1}\diamond B_{qq}\right) \diamond K^{-1} & \nonumber \\
\diamond \left(\psi^*-B_{qq}^*\diamond A_q^{-1}\diamond \psi\right) \approx & \psi^*\diamond\psi \approx |\alpha_0|^2 |g_V|^2 .
\end{align}
The expression in Eq.~(\ref{squb}) thus reads
\begin{equation}
W_{\beta}'(\alpha_0) = \frac{\exp\left( 2 |\alpha_0|^2 |g_V|^2 \right)}{1+|g_V|} .
\end{equation}

For a quantitative analysis of $|g_V|$, we use previously obtained results \cite{queez}. When the mode size of the local oscillator is much smaller than that of the pump beam, the bandwidth of the local oscillator is much larger than that of the pump beam, and thin-crystal conditions apply, the overlaps of the kernels by the mode of the local oscillator are given by
\begin{align}
\begin{split}
g_A = & \Gamma^*\diamond A\diamond\Gamma = \cosh(\Xi) = 1+g_E , \\
g_B = & \Gamma^*\diamond B\diamond\Gamma^* = \sinh(\Xi) ,
\end{split}
\end{align}
where we discarded a phase factor associated with $B$. It then follows from Eq.~(\ref{ebb}) that
\begin{equation}
|g_V|^2 \approx 2g_E-|g_B|^2 = -[\cosh(\Xi)-1]^2 \sim O(\Xi^4) .
\end{equation}
As a result, we can set $|g_V|^2=0$. The observed Wigner function for a weakly squeezed vacuum state therefore corresponds to the naive case give in Eq.~(\ref{nsqu}).

\subsection{Single-mode squeezing}

In those cases where highly squeezed states have been produced, the experimental conditions usually imply that the state represents a single mode \cite{squ15db}. When the down-conversion efficiency (squeezing parameter) is increased by strongly focussing the pump beam into the nonlinear crystal so that the Rayleigh range of the pump beam becomes comparable to the length of the crystal, the Schmidt number of the down-converted state becomes close to 1 \cite{law}. Under such conditions, the kernels of the squeezed state can be represented by
\begin{align}
\begin{split}
A(\mathbf{k}_1,\mathbf{k}_2) = & \mathbf{1}(\mathbf{k}_1,\mathbf{k}_2)
+ 2\sinh^2(\tfrac{1}{2}\Xi) \Theta(\mathbf{k}_1)\Theta^*(\mathbf{k}_2) , \\
B(\mathbf{k}_1,\mathbf{k}_2) = & \sinh(\Xi) \Theta(\mathbf{k}_1)\Theta(\mathbf{k}_2) ,
\end{split}
\end{align}
where $\Theta$ is the mode of the state.

If we assume that the mode of the state is the same as that of the local oscillator $\Theta(\mathbf{k})=\Gamma(\mathbf{k})$, then $U=V=E_{qq}=B_{qq}=0$, and the expression for the separated state in Eq.~(\ref{squ2}) would become
\begin{align}
W_{\text{sv}}[\beta](\alpha_0) = & 2\exp\left(-2|\alpha_0|^2 g_A-\alpha_0^{*2} g_B-\alpha_0^2 g_B^*\right) \nonumber \\
 & \times \mathcal{N}_0 \exp\left( -2\beta^*\diamond\beta \right) .
\end{align}
As a result, the $\beta$-dependent part is just a vacuum state, so that after tracing over $\beta$, we would recover the same expression as for the naive case given by Eq.~(\ref{nsqu}).

On the other hand, if $\Theta(\mathbf{k}) = \mu\Gamma(\mathbf{k}) + \nu\Lambda(\mathbf{k})$, where $|\mu|^2+|\nu|^2=1$, $\Gamma^*\diamond\Lambda=P\diamond\Lambda=0$ and $Q\diamond\Lambda=\Lambda$, then the coefficients in Eq.~(\ref{koefqq}) and the kernels in Eq.~(\ref{subqq}) would become
\begin{align}
\begin{split}
g_E = & 2\sinh^2(\tfrac{1}{2}\Xi) |\mu|^2 , \\
g_B = & \sinh(\Xi) \mu^2 , \\
E_{qq} = & 2\sinh^2(\tfrac{1}{2}\Xi) |\nu|^2 \Lambda\Lambda^* , \\
B_{qq} = & \sinh(\Xi) \nu^2 \Lambda\Lambda .
\end{split}
\end{align}
Moreover,
\begin{align}
\begin{split}
E\diamond\Gamma = &  2\sinh^2(\tfrac{1}{2}\Xi) \left( |\mu|^2\Gamma + \mu^*\nu\Lambda \right) , \\
B\diamond\Gamma^* = & \sinh(\Xi) \left( \mu^2\Gamma + \nu\mu\Lambda \right) .
\end{split}
\end{align}
Hence,
\begin{align}
\begin{split}
U = & 2\sinh^2(\tfrac{1}{2}\Xi) \nu\mu^*\Lambda , \\
V = & \sinh(\Xi) \nu\mu\Lambda , \\
\psi = & \left[2\sinh^2(\tfrac{1}{2}\Xi) \mu^*\alpha_0+\sinh(\Xi) \mu\alpha_0^*\right]\nu\Lambda .
\end{split}
\end{align}

\begin{figure}[ht]
\centerline{\includegraphics{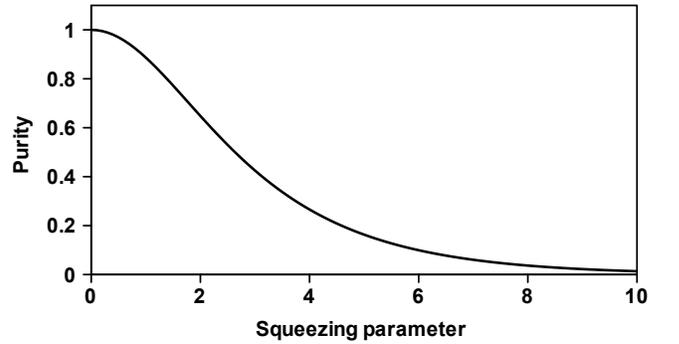}}
\caption{Purity of the observed single-mode squeezed vacuum state as a function of the squeezing parameter for $|\mu|=\tfrac{1}{2}$.}
\label{purity}
\end{figure}

With the aid of these quantities and the expressions in Eqs.~(\ref{distor}) and (\ref{squ3}), we can determine the expression for the observed Wigner function. It reads
\begin{align}
W_{\text{sv}}(\alpha) = & \frac{2}{\sqrt{1+4|\mu|^2|\nu|^2\sinh^2(\tfrac{1}{2}\Xi)}} \nonumber \\
 & \times \exp\left[-\frac{2|\alpha|^2
 +4|\alpha|^2|\mu|^2\sinh^2(\tfrac{1}{2}\Xi)}{1+4|\mu|^2|\nu|^2\sinh^2(\tfrac{1}{2}\Xi)}\right. \nonumber \\
& \left. -\frac{\alpha^{*2}\mu^2\sinh(\Xi)+\alpha^2\mu^{*2}\sinh(\Xi)}{1+4|\mu|^2|\nu|^2\sinh^2(\tfrac{1}{2}\Xi)}\right] .
\label{emsv}
\end{align}
For $\mu=1$, the expression becomes equivalent to Eq.~(\ref{nsqu}), and for $\mu=0$, it becomes that of a vacuum state.

In general Eq.~(\ref{emsv}) represents a mixed state, with
\begin{equation}
\text{purity} = \left[1+4|\mu|^2|\nu|^2\sinh^2(\tfrac{1}{2}\Xi)\right]^{-1/2} .
\end{equation}
The largest amount of mixing is obtained for $|\mu|^2=\tfrac{1}{2}$. The purity for this case is plotted in Fig.~\ref{purity} as a function of the squeezing parameter.

\begin{figure}[ht]
\centerline{\includegraphics{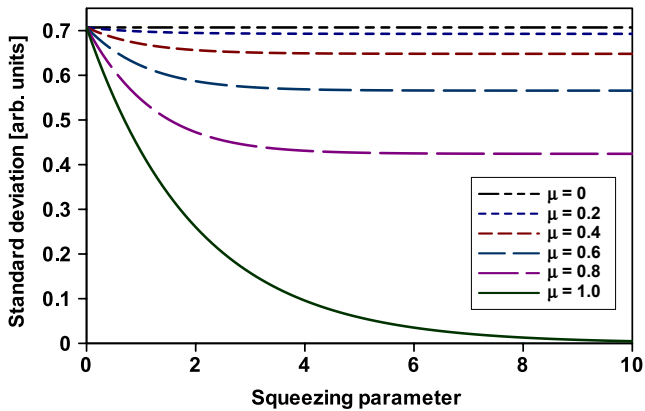}}
\caption{Minimum standard deviation of the observed single-mode squeezed vacuum state as a function of the squeezing parameter for different values of $|\mu|$.}
\label{std}
\end{figure}

The amount of squeezing is also diminished by $\mu$. Along the squeezed direction, the standard deviation is
\begin{equation}
\sigma_{\text{min}} = \frac{1}{\sqrt{2}} \left[1-|\mu|^2+|\mu|^2\exp(-\Xi)\right]^{1/2} .
\end{equation}
The standard deviation along the squeezed direction is plotted in Fig.~\ref{std} as a function of the squeezing parameter for different values of $|\mu|$.

\section{Conclusions}

Several observations follow from the analyses provided above. In general, we see that, unless the input state is parameterized by a single parameter function and both the modes of the local oscillation and the detection system match this parameter function exactly, which assumes {\em a priori} knowledge of the state's parameter function, the homodyning tomography process produces observed Wigner functions that are distorted. These distortions are partly determined by the experimental conditions and partly by the nature of the state that is being measured. Here we summarize the salient features of these distortions.

The main experimental conditions that influence the distortions are those associated with the local oscillator and the detection process. The local oscillator is usually parameterized by a single mode, which determines the spatiotemporal properties of the observed Wigner function. The rest of the spatiotemporal degrees of freedom of the input state are traced out and this trace process can affect the observed Wigner function. The optical power of the local oscillator plays an important role in the process. It sets a boundary for the charateristic function of the state outside of which the charateristic function is set equal to zero. Unless the charateristic function lies inside the boundary, it would be distorted due to clipping. On the phase space, the power (or average number of photons) of the local oscillator determines the resolution of the observed Wigner function. More powerful local oscillators produce better resolution. If the average number of photons in the local oscillator is comparable to those of the state being measured, the resolution would be on the order of the minimum uncertainty area. The effect of the finite resolution is a broadening of the observed Wigner function, which implies that it is rendered as a mixed state.

Provided that the efficiency of the detection process is the same for all photons, regardless of their spatiotemporal degrees of freedom, it only causes a global scaling of the observed Wigner function. This scaling effect can be readily removed by rescaling the phase space coordinates. In those cases where the detection efficiency depends on the spatiotemporal degrees of freedom of the photons, such as would be determined by the overlap with the mode of a single-mode detector, it contributes to the distortion of the observed Wigner function. Since, the homodyne tomography process does not measure the state directly, but instead measures a cross-correlation distribution from which the observed Wigner function is computed, the efficiency does not appear as a probability in the mixture. Instead, our analysis shows that it produces a scaling of the coordinates.

Nevertheless, some distortions are associated with the loss of purity in the observed Wigner function, even if the state that is being measured is pure. There are different mechanisms responsible for this effect. For a displaced state, such as a coherent state, the observed Wigner function after scaling corrections generally has an increased width, representing a loss of purity. This increase in width is caused by the intrinsic kernel function of the homodyning process. It is proportional to the average number of photons in the state and inversely proportional to the average number of photons in the local oscillator. Therefore, a local oscillator with a larger optical power will produce an observed Wigner function with a better purity. When the state is located at the origin and is not displaced, the contribution to the loss of purity due to the intrinsic kernel function of the homodyning process is negligible for a suitably large average number of photons in the local oscillator, with the possible exception of severely squeezed states.

However, there are other ways in which states that are located at the origin can lose purity. These cases are related to the properties of the states themselves and result from the trace that removes the degrees of freedom not related to those of the local oscillator and the detection system. If the state is not parameterized by a single parameter function, such as squeezed states, or if its parameter function does not match the mode functions of the local oscillator and the detection system, then the trace causes contributions to the observed Wigner function that distort it and contribute to a loss of purity. The reason can be found in the fact that the spatiotemporal degrees of freedom that are associated with the mode of the local oscillator and the detection system could be entangled with those that are traced out. As a result, the observed Wigner function becomes that of a mixed state. The distortions can also take on other forms. For instance, in the case of a squeezed state, it can reduce to amount of squeezing in the state.

The Wigner functional analysis of the homodyning tomography process reveals an important aspect of quantum optical states. The marginal distributions that are obtained by integrating the observed Wigner function along one direction are always non-negative. It indicates that the homodyning process always produces observed Wigner functions with valid marginal distributions. However, the input state is represented by a Wigner functional on an infinite-dimensional functional phase space. As a result, the observed Wigner function requires that all the unobserved spatiotemporal degrees of freedom are traced out. This process plays an important role in those cases where the Wigner functional is negative in some regions, such as Fock states and photon-subtracted or -added states \cite{trepstheorem,lvovsky}. In a practical scenario, the parameter function that parameterizes a state would not be known before hand, and it would therefore not be possible to match it to the mode of the local oscillator and the detection system. Without the contribution of the trace over the unobserved spatiotemporal degrees of freedom, these negative regions would not be filled up when the marginal distributions are computed from the observed Wigner function. Therefore, in such practical cases, the trace process may affect those parts of the Wigner functional that become part of the observed Wigner function --- those degrees of freedom that are traced out may contribute to the observed Wigner function and are not simply discarded.

\section*{Acknowledgement}

This work was supported in part by funding from the National Research Foundation of South Africa (Grant Number: 118532) and from the Department of Science and Innovation (DSI) through the South African Quantum Technology Initiative.

\end{document}